\documentclass[reprint,superscriptaddress,
 amsmath,amssymb,
prl,
]{revtex4-2}

\usepackage{dcolumn}
\usepackage{bm}

\usepackage{hyperref}
\usepackage{graphicx}
\usepackage{subfiles}
\usepackage{epstopdf,epsfig,subfloat,subfig,floatrow}

\begin{document}

\preprint{APS/123-QED}

\title{Small Rarefaction, Large Consequences: Limits of Navier–Stokes Turbulence Simulations}

\author{Songyan Tian}
\author{Lei Wu}

\affiliation{Department of Mechanics and Aerospace Engineering, Southern University of Science and Technology, Shenzhen 518055, China}

\date{\today}

\begin{abstract}
We conduct numerical simulations of rocket plume impingement on a lunar landing surface using two complementary frameworks: the Boltzmann equation, which naturally captures rarefied gas dynamics, and the  Navier–Stokes (NS) equations, the conventional workhorse for turbulent flow simulations. We show that subtle rarefaction effects—long considered negligible in turbulent regimes—can become locally dominant within shear layers where viscous stresses predicted by the NS constitutive relation undergo sign reversals. This phenomenon, which we term constitutive degeneracy, produces order-one relative errors in predicted surface shear stress and heat flux. Our results demonstrate that turbulence can expose hidden limits of NS equations with broad implications for high-speed aerodynamics and planetary exploration.
\end{abstract}

\maketitle



\textbf{Introduction}---The NS equations form the foundation of most theoretical and computational descriptions of fluid flow, from aerodynamics and combustion to geophysical and astrophysical turbulence. Their validity in gas dynamics relies on the assumption that the Knudsen number (Kn, the ratio of the molecular mean free path to a characteristic flow length) is sufficiently small. When this condition is violated, rarefaction (non-equilibrium) effects undermine the Newtonian stress–strain relationship, causing the NS equations to lose accuracy. By contrast, the Boltzmann equation provides a kinetic description that remains valid for dilute gases across all flow regimes \cite{Chapman1990Mathematical}, encompassing turbulent and laminar flows from the continuum limit through the slip and transition regimes to the free-molecular regime as Kn increases.

In turbulent flows, the characteristic length scale relevant to molecular transport is often taken to be the Kolmogorov scale. Classical turbulence theory has long recognized that, when this scale is adopted as the characteristic flow length, even flows with large global Reynolds numbers may formally admit locally non-negligible Kn \cite{Tennekes1972first}. This estimation has motivated sustained efforts to identify signatures of molecular rarefaction within turbulent flows.

Despite these expectations, decades of research have produced little evidence that rarefaction effects play a significant role in bulk turbulent flows \cite{Stefanov1992Monte, Stefanov1993Monte, Stefanov2000Monte, Komatsu2014glimpse, Gallis2017Molecular, Li2018Thermal}. Direct numerical studies based on kinetic descriptions—including direct simulation Monte Carlo solutions of the Boltzmann equation~\cite{Bird1994Molecular}—have shown that canonical turbulence statistics, such as the Kolmogorov energy spectrum, remain essentially unchanged even when the smallest turbulent scales approach the molecular mean free path \cite{Gallis2017Molecular}. When deviations from NS predictions are observed, they are confined to very high wavenumbers and can be attributed primarily to thermal fluctuations rather than genuine rarefaction effects \cite{Gallis2022PRL, Bell2022ThermalFluctuations}. These findings have reinforced the prevailing view that, within homogeneous turbulence, rarefaction effects remain dynamically passive and do not influence macroscopic turbulent behavior. Consequently, the NS equation remains reliable for turbulence simulations.

However, real-world turbulence is seldom homogeneous. Strong spatial inhomogeneities, such as shock waves, rapid expansions, and intense shear layers, can create localized environments in which continuum assumptions are weakened, even though the flow remains turbulent at larger scales. In such settings, turbulence and molecular rarefaction need not be cleanly separated in scale or space; instead, they may coexist and interact within the same flow field. Despite this possibility, most existing studies have treated turbulence and rarefaction as essentially independent phenomena \cite{Mahesh2013Interaction, Morris2016Lunar}, leaving unresolved whether weak rarefaction effects can greatly influence macroscopic quantities when both are simultaneously present.

Here, we address this question using a canonical yet technologically consequential configuration: turbulent rocket plume impingement during planetary landing~\cite{Kim2025survey}. In this setting, surface stress and heat flux are the primary quantities of engineering relevance, governing material loading, thermal gradients, and long-term degradation under repeated plume exposure. By  time-resolved simulations of the Boltzmann equation, we demonstrate that weak rarefaction effects can exert a decisive influence in bulk turbulence, leading to substantial deviations in surface stress and heat flux that are not captured by the NS equations.

\begin{figure*}[t]
    \centering
    \sidesubfloat[]{\includegraphics[width=0.45\linewidth]{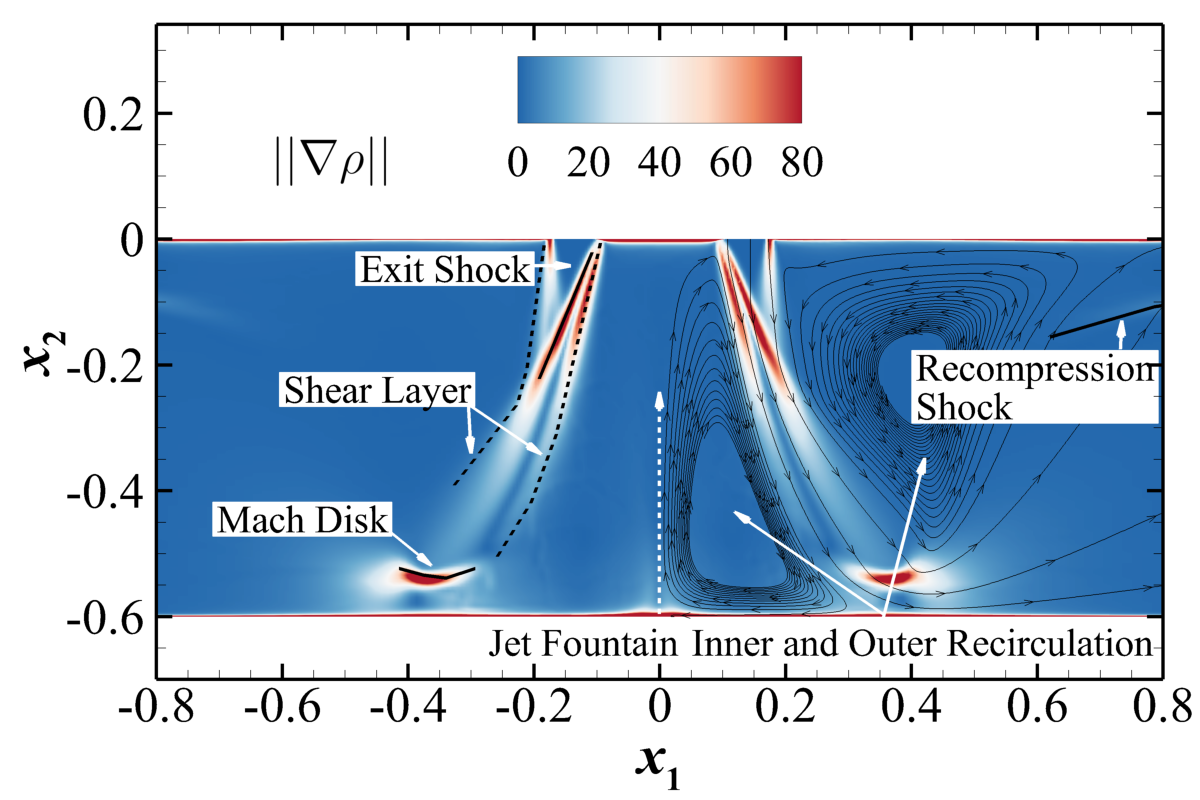}} 
    \sidesubfloat[]{\includegraphics[width=0.45\linewidth]{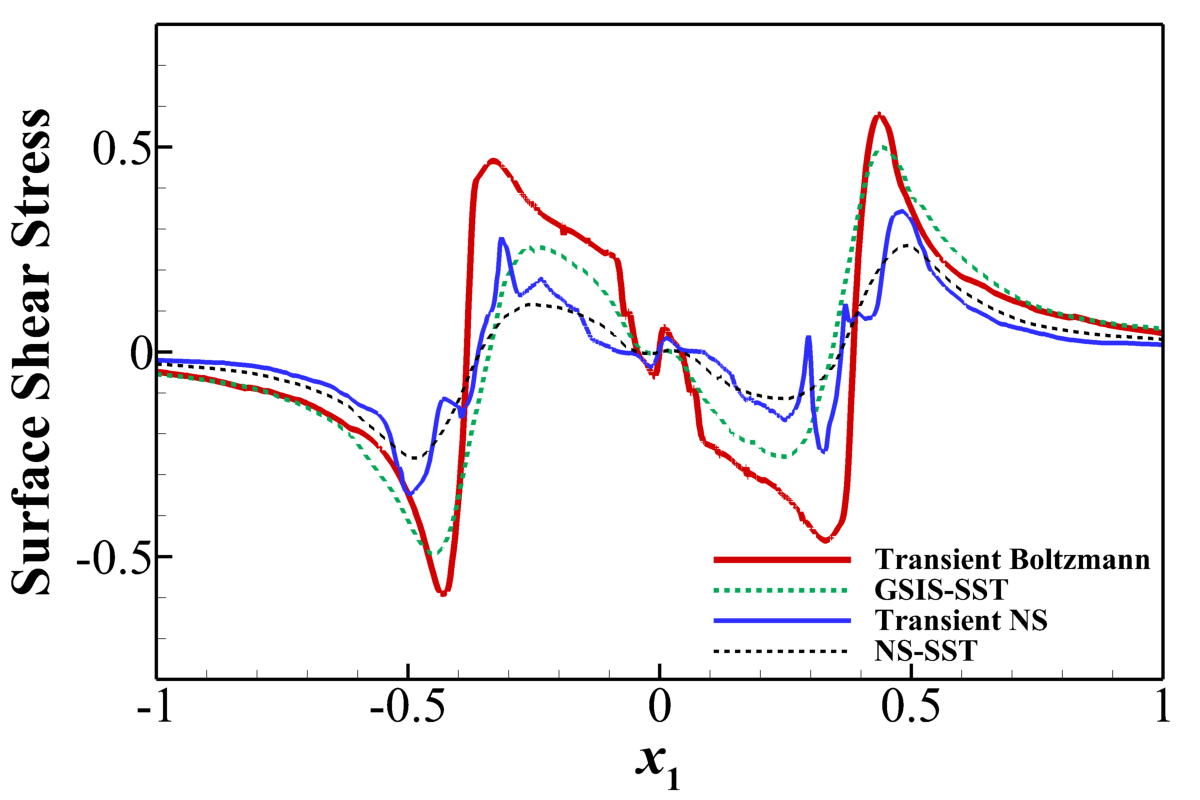}}\\
    \sidesubfloat[]{\includegraphics[width=0.45\linewidth]{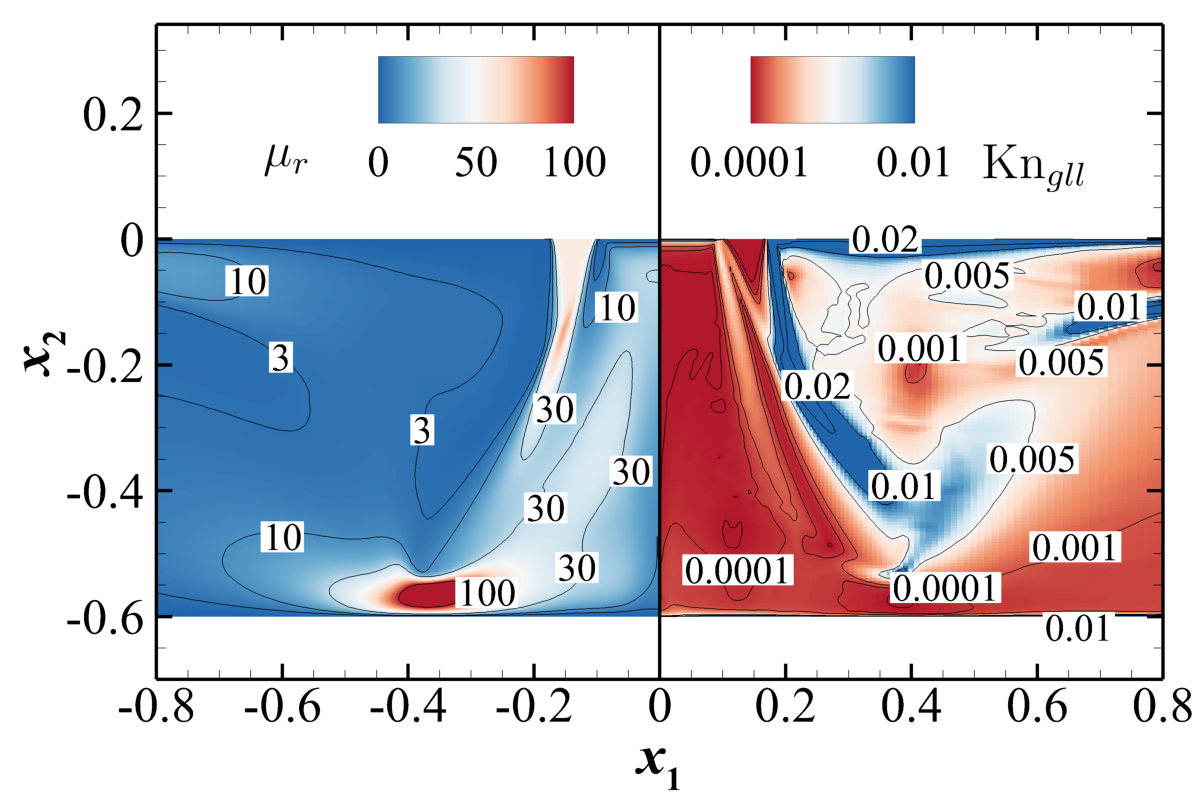}}
    \sidesubfloat[]{\includegraphics[width=0.45\linewidth]{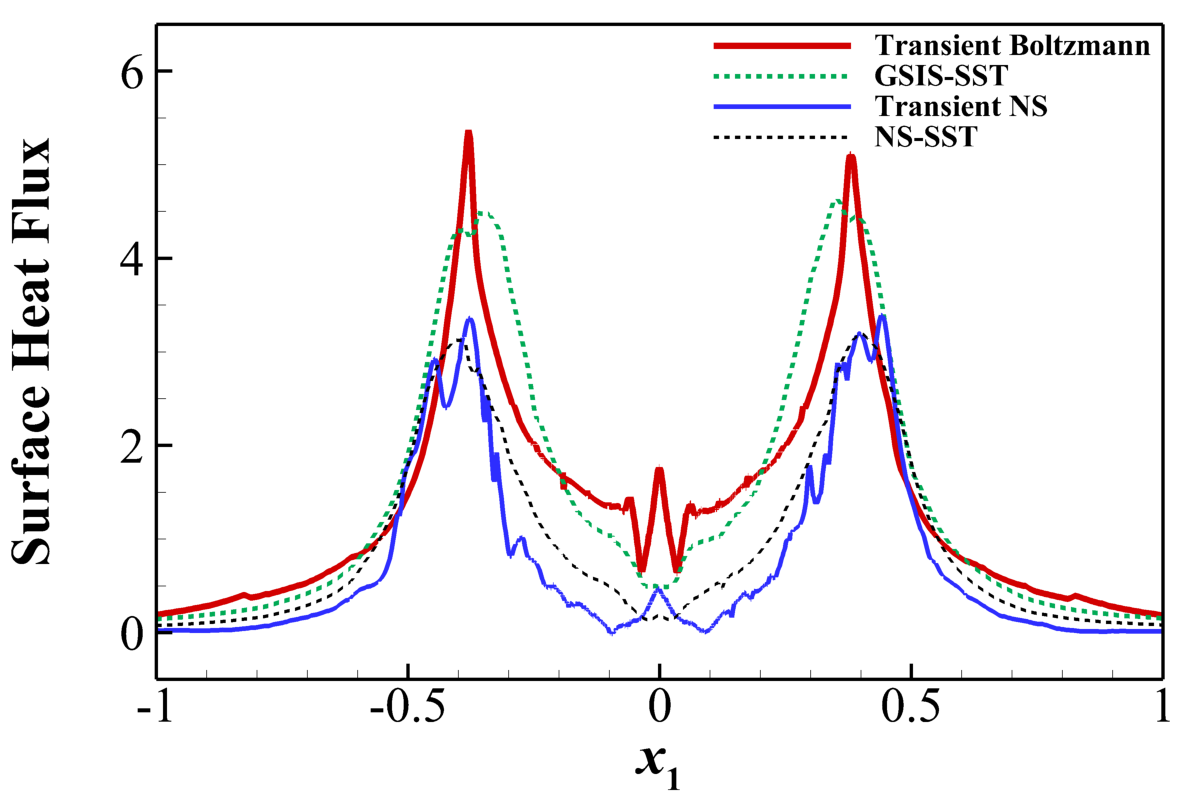}}\\
    \sidesubfloat[]{\includegraphics[width=0.45\linewidth]{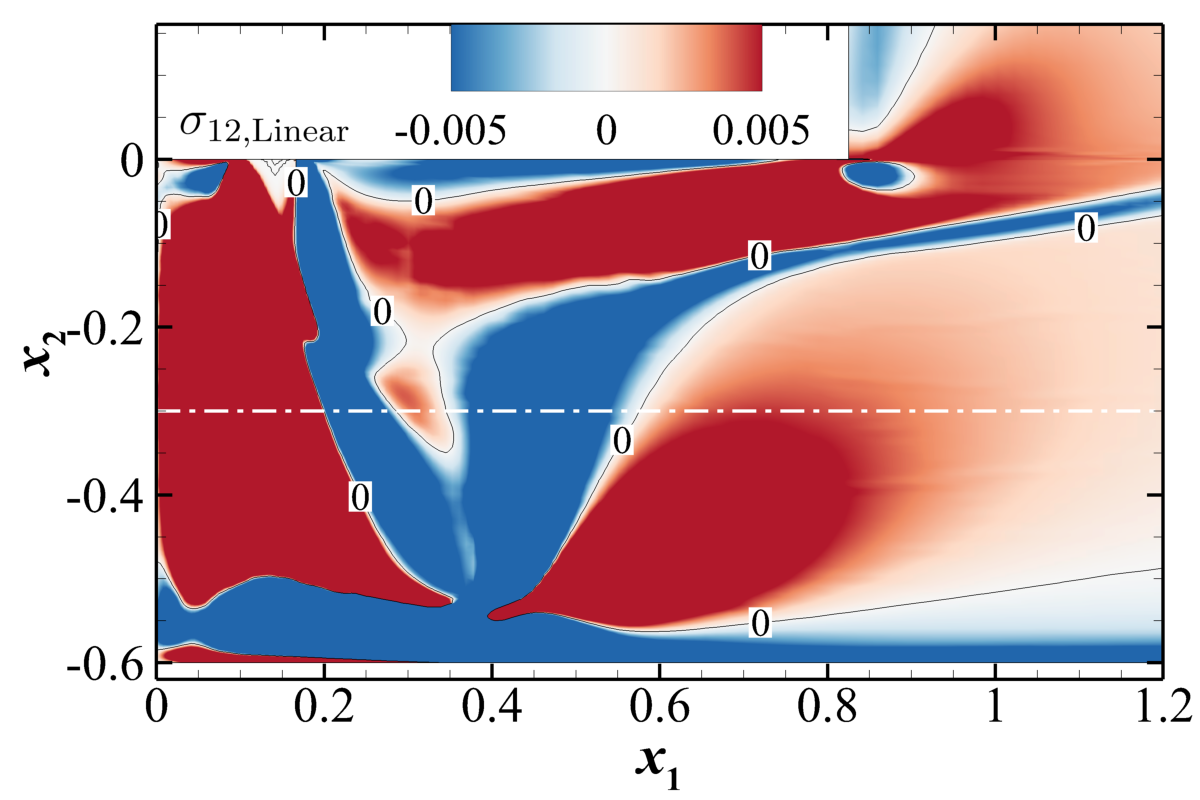}}
    \sidesubfloat[]{\includegraphics[width=0.45\linewidth]{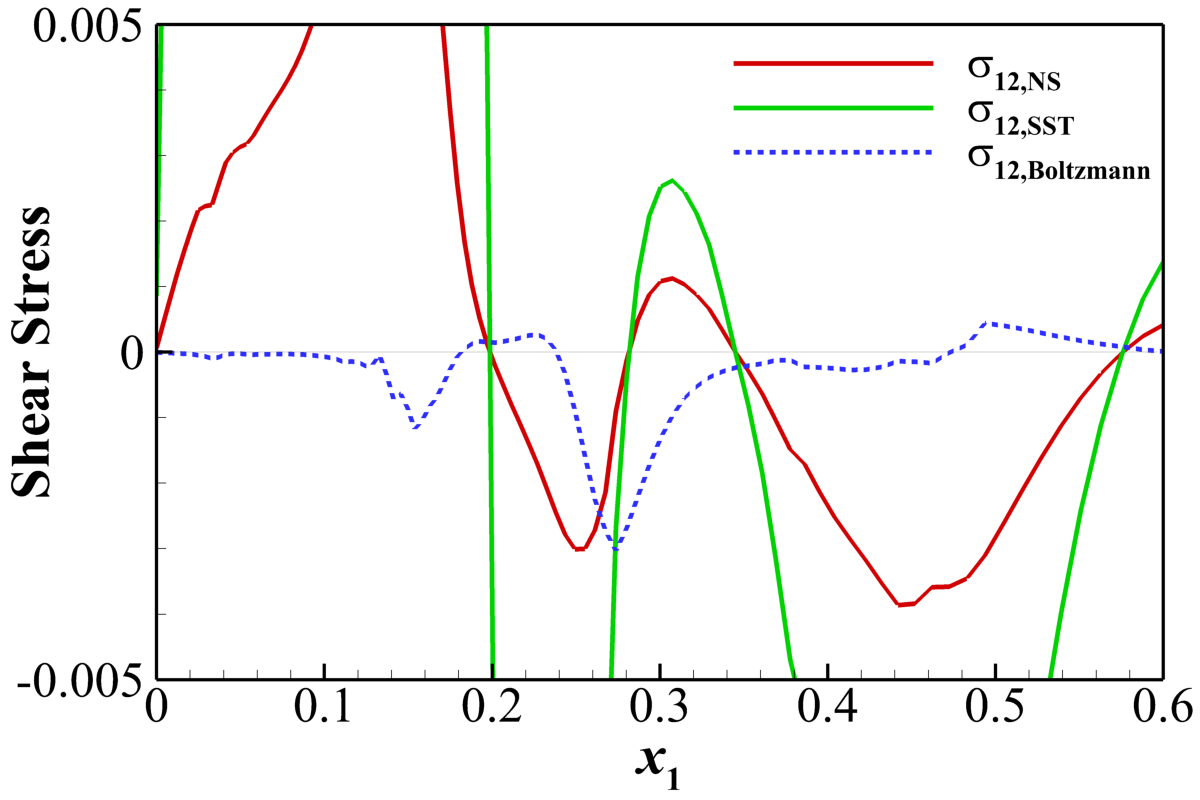}}\\
    \caption{ 
    (a) The density gradient and streamlines. 
    (b) Shear stress at the ground surface. 
    (c) The turbulent-to-laminar viscosity ratio $\mu_r$ and the local Knudsen number $\text{Kn}_{gll}$, in the left and right half-domains, respectively. 
    (d) Heat flux at the ground surface. 
    (e) Contour of the NS shear stress $\sigma_\text{12,Linear}=\sigma_\text{12,NS}+\sigma_\text{12,SST}$. 
    (f) The laminar, turbulent, and non-equilibrium stresses at $x_2=-0.3$~m. 
    }
    \label{fig:intro}
\end{figure*}

\textbf{Problem setting}---As shown in Fig.~\ref{fig:intro}(a), we consider a two-dimensional lunar landing module (LLM) with a width of 1.65~m and a height of 0.6~m. Two identical rocket engines, each with a nozzle exit diameter of $L = 0.08$m, are mounted symmetrically at $x_1 = \pm 0.135$~m and $x_2=0$~m.
Flow conditions of the nitrogen exhaust gas at the nozzle exit are characterized by a Mach number of 5.45, a static pressure of $P=172.5$~Pa and a temperature of $T=400$~K, corresponding to a Knudsen number of $\text{Kn}=\frac{\lambda}{L}=\frac{\mu}{P L} \sqrt{\frac{\pi RT}{2}}=6.87\times10^{-4}$ and a Reynolds number of $\text{Re}=\frac{P L Ma}{\mu} \sqrt{\frac{7}{5RT}}=1.18\times10^{4}$, where $\lambda$ and $\mu$ are molecular mean free path and physical viscosity, respectively, and $R$ is the gas constant of nitrogen. The rocket jets are assumed to have a turbulence intensity of 5\%, and the jet diameter is assigned as the turbulent length scale.
The whole rectangular computational domain has a width of 20~m and a height of 3.6~m, where the top boundary at $x_2=3$~m and the two vertical boundaries at $x_1=\pm10$~m are vacuum. The landing surface with a temperature of 295~K is located at $x_2=-0.6$~m.

We adopt the modified Rykov kinetic model to describe the gas dynamics~\citep{Li2021Uncertainty}:
\begin{equation}\label{Bol_eq}
    \frac{\partial f}{\partial t}+\bm{v}\cdot\frac{\partial f}{\partial \bm{x}}=Q,
\end{equation}
where $f(t,\bm{x},\bm{v})$ is the velocity distribution function, $\bm{x}=(x_1,x_2,x_3)$ is the spatial coordinate, $\bm{v}=(v_1,v_2,v_3)$ is the molecular velocity, and $Q$ is the simplified Boltzmann collision operator. 
The kinetic model is solved using the discrete velocity method \citep{Chu1965a,Aristov2001Direct, Zhang2024Efficient}, in which both physical and molecular velocity spaces are discretized. Here, the molecular velocity space is reduced and truncated to $v_1 \in (-20,20) \sqrt{RT}$~m/s and $v_2 \in (-24,24)\sqrt{RT}$~m/s, which is discretized uniformly into 7,680 cells.

\textbf{Coarse-grained simulation}---Although the Boltzmann equation encapsulates the gas dynamics across turbulent to rarefied regimes, the computational demands of turbulence simulation are overwhelming, since in regions of high Reynolds number, the spatial grid must be finely resolved. We first employ a coarse-grained multiscale framework that couples a mesoscopic kinetic solver with a macroscopic Reynolds-averaged turbulence closure, in a coarse spatial grid of 101,162 nonuniform cells refined beneath the LLM. This framework, termed GSIS–SST \citep{Tian2025Multiscale}, combines the general synthetic iterative scheme (GSIS)—which enables fast-converging, asymptotic-preserving simulation of the rarefied gas flows \citep{Su2020Can, Su2020Fast}—with the shear stress transport (SST) turbulence model to represent turbulent dynamics in the continuum regime. The key ingredient in GSIS-SST is that the constitutive relations consist of three contributions, i.e., the laminar and turbulent stress ($\bm{\sigma}_\text{NS}$ and $\bm{\sigma}_\text{SST}$) described by Newton’s law with the physical and turbulent viscosities ($\mu$ and $\mu_\text{SST}$), and the high-order (non-equilibrium) stress $\bm{\sigma}_\text{Boltzmann}$ extracted from the Boltzmann equation \eqref{Bol_eq}:
\begin{equation}\label{sigma_3components}
\begin{aligned}
\bm{\sigma}= \underbrace{\bm{\sigma}_\text{NS}}_{-2\mu\bm{S} }
+\underbrace{\bm{\sigma}_\text{SST}}_{-2\mu_\text{SST}\bm{S}}
+\bm{\sigma}_\text{Boltzmann}, 
\end{aligned}
\end{equation}
where $\bm{S}=\frac{1}{2}(\nabla \bm{u}+\nabla \bm{u}^\top-\frac{2}{3}\nabla \cdot \bm{u} I)$ is the traceless strain-rate tensor and $\bm{u}$ is the flow velocity; the same holds for the heat flux. 
It can be inferred from the Chapman–Enskog expansion \citep{Su2020Fast} that, when Re is large and Kn is small, the higher-order stress scales as \(\text{Kn}^2\); consequently, the GSIS–SST solver degenerates to the NS–SST model. In contrast, when Re is small and Kn is large, the SST-modeled turbulent stress rapidly dissipates, and the GSIS–SST solver reduces to a Boltzmann solver. In the intermediate regimes, both turbulence and rarefaction effects are captured in a coarse-grained manner. Unlike conventional approaches that rely on a priori domain decomposition into continuum and rarefied regions, the present framework autonomously distinguishes between flow regimes during the simulation.

Figure~\ref{fig:intro}(a) presents the GSIS-SST solution for density gradient magnitude $||\nabla\rho||$. As the rocket jets enter the domain, they undergo rapid outward deflection through exit shocks and continue to accelerate toward the ground. Upon impingement, Mach disks form and decelerate the jets, followed by renewed expansion as the flow spreads into the surrounding low-density environment. A fraction of the jet flow reverses direction and impinges on the lower surface of LLM, generating additional shock structures (the recompression shock) near the LLM lower corners.
In the vicinity of jets, distinct recirculation regions develop. The inner recirculation zone, confined between the two jets, is characterized by relatively high density and a small molecular mean free path. It contains two large vortices with a central fountain-like jet, a hallmark of closely spaced multi-jet impingement systems~\citep{Zuckerman2006Jet}. The elevated pressure in this inner zone forces the jets to deflect immediately upon exiting the nozzles. In contrast, the outer recirculation zones occupy the region between the jets and the lower surface of the LLM and open toward the near-vacuum environment. Owing to their low density and pressure, these zones promote continued outward expansion of the jets and enhance their lateral deflection.

Figure~\ref{fig:intro}(b,d) shows the ground distributions of the shear stress and heat flux predicted by the GSIS-SST, as well as those from the NS solver with SST turbulence closure (NS-SST).
Across the surface, both quantities exhibit robust, symmetric multi-peak structures characteristic of jet impingement. However, 
NS-SST underpredicts the magnitudes of key shear-stress extrema by approximately 25$\sim$30\% relative to GSIS-SST, with the largest differences occurring near dominant recirculation and impingement regions. A similar underprediction is observed in the surface heat flux: the NS-SST model yields peak values that are approximately 50\% lower and exhibits a more rapid downstream decay compared with the multiscale solution.

These discrepancies are not confined to isolated points but persist across regions that dominate the surface's stress and heat load. 
Notice that the turbulent-to-laminar viscosity ratio in Fig.~\ref{fig:intro}(c) implies that the whole areas between the jets should be turbulence prevailing, especially around the impingement locations.
The surface response thus indicate that, even in a nominally turbulent regime, continuum-based turbulence modeling can substantially underestimate engineering-relevant surface stresses and heat transfer.

\textbf{Degeneracy of the NS constitutive relation}---
To identify the underlying rarefaction effects in the jet-controlled regions, we first plot in Fig.~\ref{fig:intro}(c) the gradient-length local Knudsen number~\citep{Boyd1995Predicting}, 
\begin{equation}
    \text{Kn}_{gll}=\frac{\lambda}{\rho} |\nabla \rho|.
\end{equation}
It is observed that the flow remains predominantly in the continuum regime, only marginally entering the slip-flow region (where $\text{Kn}_{gll}\approx0.02$) and staying well below the strongly rarefied regime.

\begin{figure*}[t]
    \centering
    \sidesubfloat[]{\includegraphics[width=0.45\linewidth]{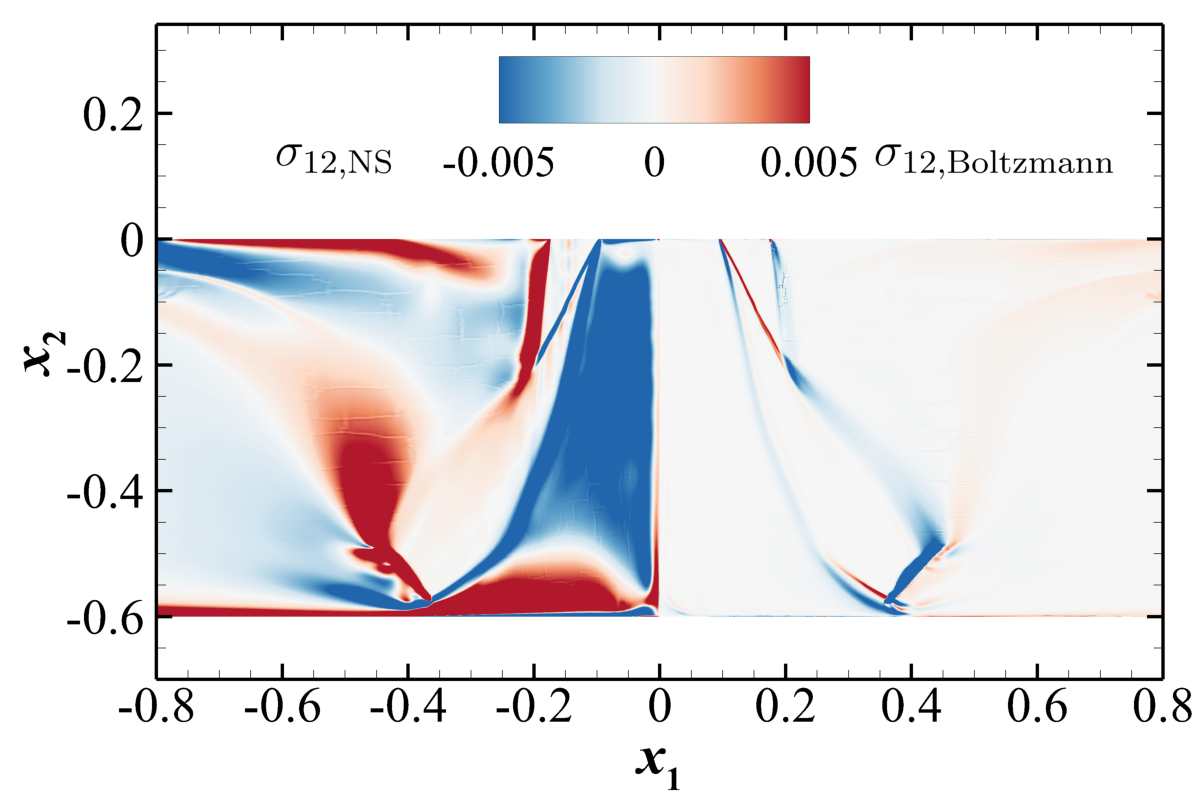}} 
    \sidesubfloat[]{\includegraphics[width=0.45\linewidth]{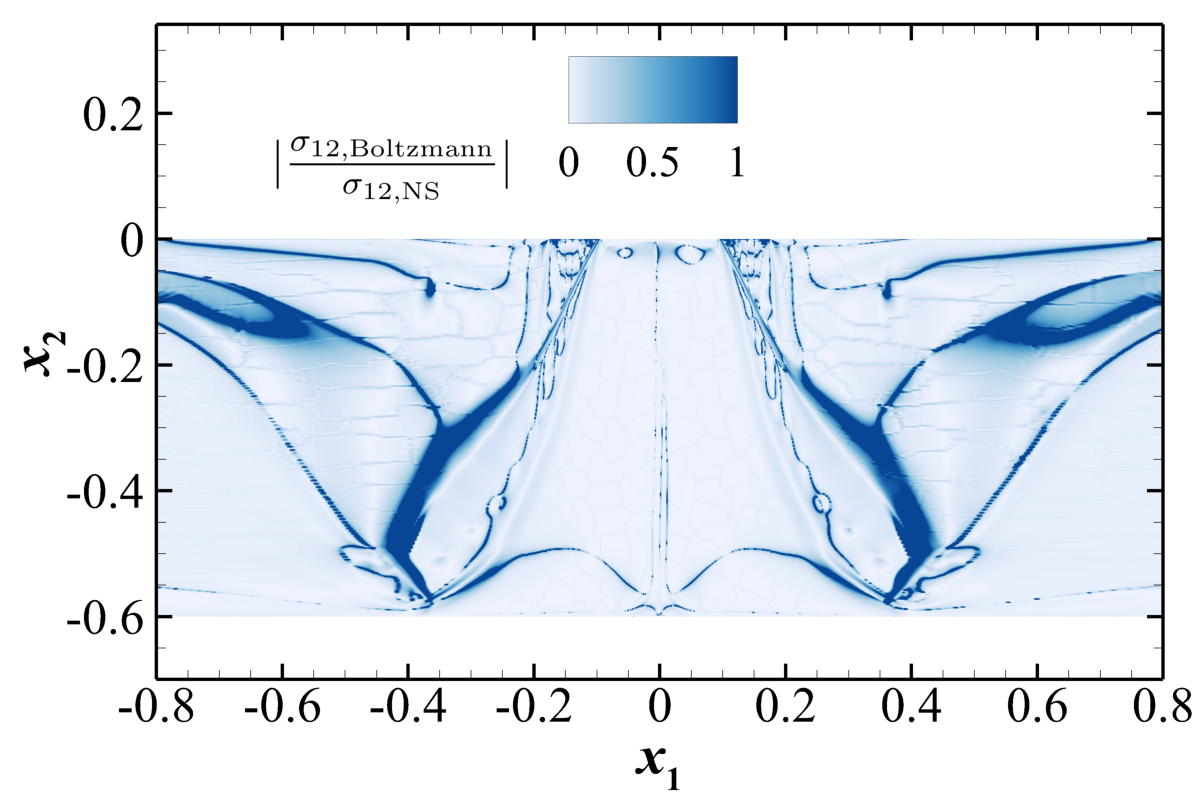}}\\
    \sidesubfloat[]{\includegraphics[width=0.45\linewidth]{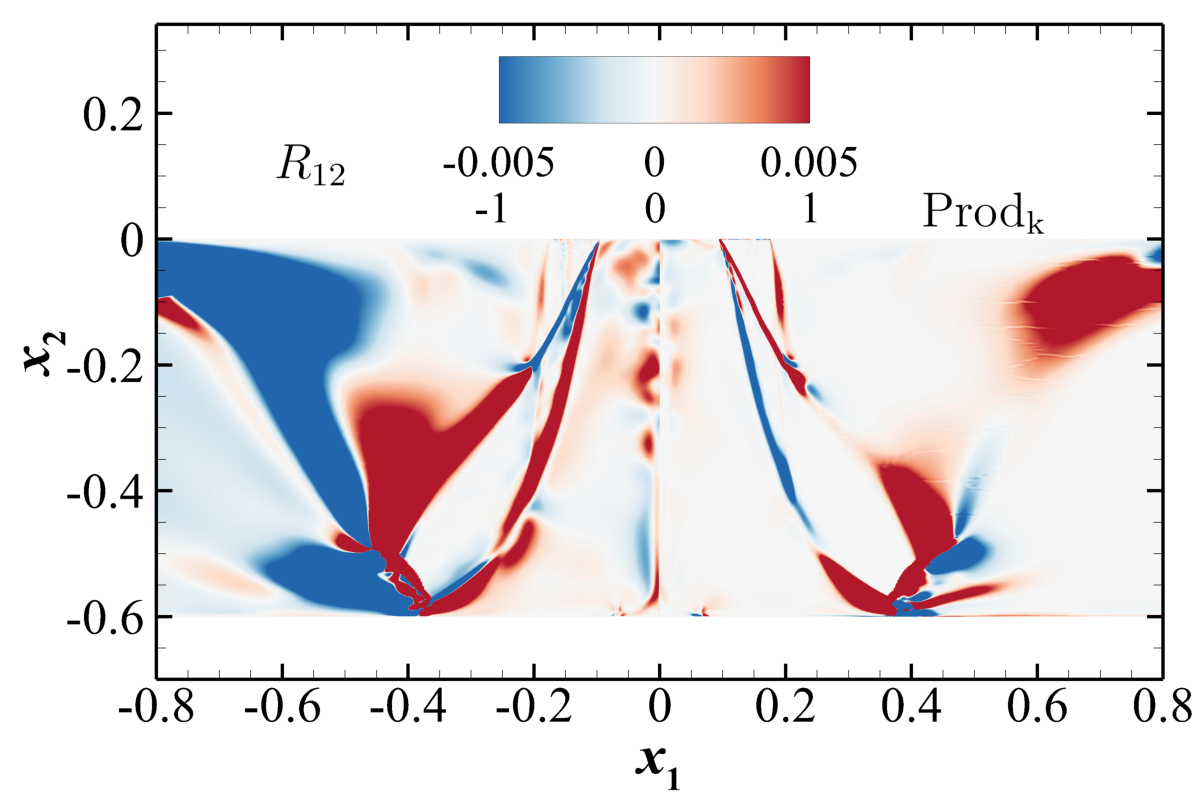}}
    \sidesubfloat[]{\includegraphics[width=0.45\linewidth]{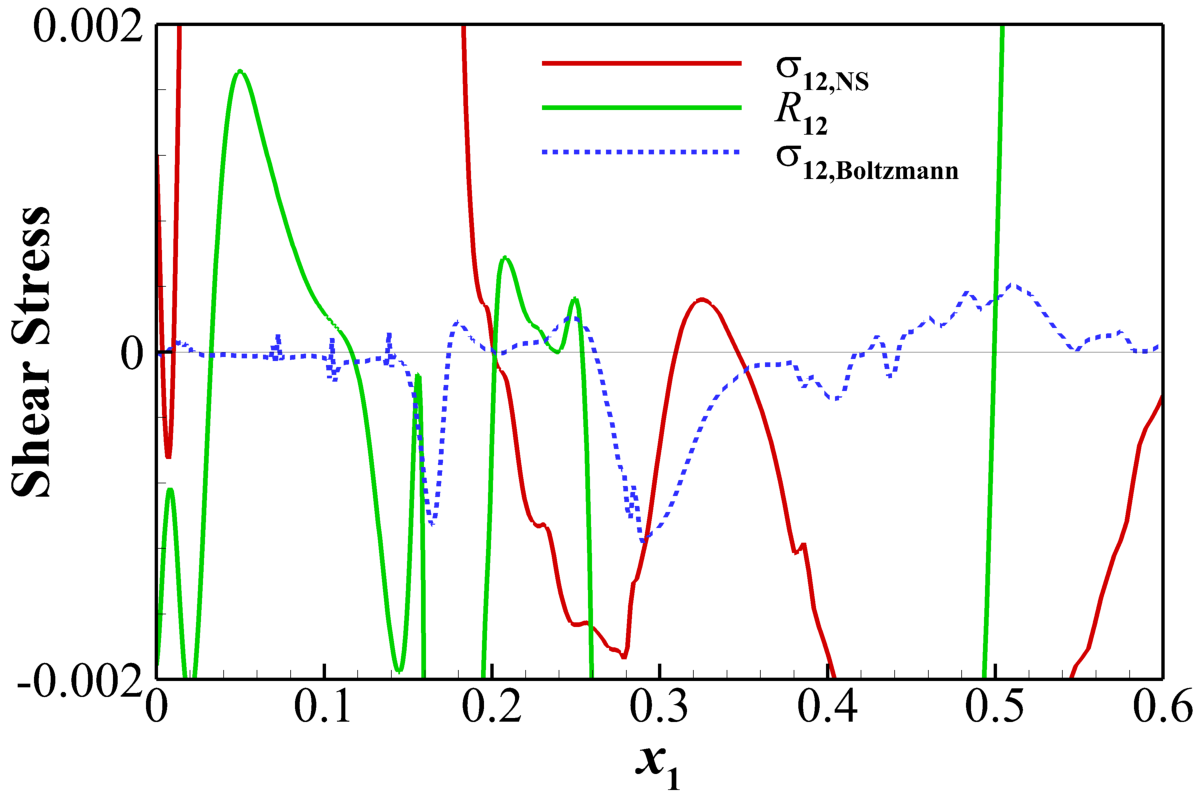}}
    \caption{ Direct numerical simulation of the Boltzmann equation using the transient GSIS solver in turbulent-model-free mode~\cite{Zeng2023GSIS}.
    (a,b) The NS and Boltzmann shear stress, and their relative strength. 
    (c) The Reynolds shear stress $R_{12}$ and the turbulence production term $\text{Prod}_\text{k}$.  
    (d) The laminar, Reynolds, and non-equilibrium stresses at $x_2=-0.3$~m. 
    }\label{fig:intro2}
\end{figure*}

We then plot in Fig.~\ref{fig:intro}(e) the distribution of the linear NS shear stress $\sigma_\text{12,Linear}$, which is the sum of laminar and turbulent contributions. The zero-stress contour lines ($\sigma_\text{12,Linear}=0$) are  highlighted. The vanishing of the NS shear stress does not imply small velocity gradients, but occurs when $\partial u_1/\partial x_2$ and $\partial u_2/\partial x_1$ attain comparable magnitude but opposite sign, a condition that occurs naturally in regions dominated by flow turning, pure compression/expansion, or recirculation.
Under such kinematic configurations, the local deformation is governed by rotational or extensional modes rather than shear, causing the Newtonian shear stress (obtained from the first-order Chapman–Enskog expansion of the Boltzmann equation) to collapse along extended spatial paths despite sustained velocity gradients. However, higher-order non-equilibrium stresses do not vanish. 
For example, Fig.~\ref{fig:intro}(f) shows that, around $x_1\approx0.2,0.28,0.34,0.58$~m, the NS stress undergoes sign reversals; 
the higher-order rarefaction contribution $\sigma_\text{12,Boltzmann}$, although smaller in absolute magnitude, becomes locally dominant around $x_1\in(0.25,0.35)$~m.

Thus, this finding challenges the conventional view that rarefaction effects become significant only when their absolute magnitude is large. Instead, the decisive mechanism is the local collapse of the NS continuum stress, which disrupts the regular perturbative ordering of the constitutive relation in the Chapman-Enskog expansion. When such degeneracy occurs, even weak non-equilibrium corrections can locally dominate the stress balance and are subsequently propagated and nonlinearly amplified by the evolving flow.

\textbf{Direct simulations of the Boltzmann equation}---Since the SST model is a coarse-grained turbulence model, its applicability within the GSIS–SST solver may be questioned. To exclude any influence introduced by turbulence modeling, we perform the time-resolved simulations by directly solving the Boltzmann equation \citep{Zeng2023GSIS}.
Grid-independence studies confirm that the final mesh ensures a minimum spatial spacing below the LLM of approximately twice the local Kolmogorov length scale, resulting in a nonuniform spatial mesh of 957,204 cells.
Time integration is carried out using a second-order backward Euler scheme with an implicit dual-time stepping strategy, where the time step is $0.43L/\sqrt{RT}=0.0001$ second. To initiate unsteady jet dynamics, synthetic velocity perturbations are imposed at each nozzle exit, constructed from 100 superimposed eddies with randomized spatial locations, characteristic frequencies, phases, and amplitudes~\citep{Kraichnan1970Diffusion,Davidson2006Hybrid}. The perturbation field is uniformly scaled to achieve a nominal turbulence intensity of 5\%. The initial condition is taken from the steady GSIS-SST solution to reduce computational cost.

The ground shear stress and heat flux are time averaged over $t\in(4308, 8616)L/\sqrt{RT}=(1.0,2.0)$ seconds
after discarding initial transients. 
Figure~\ref{fig:intro2}(a) shows that the time-averaged flow field is still predominantly governed by the NS stress contribution, whereas the Boltzmann higher-order stress displays a spatially localized distribution that closely resembles that obtained in the steady GSIS-SST solution. Figure~\ref{fig:intro2}(b) shows that, while $\sigma_\text{12,Boltzmann}$ is smaller in magnitude, it becomes dominant in regions where $\sigma_\text{12,NS}$ is close to zero. 

In addition, the resolved unsteady momentum flux, quantified by the Favre-averaged correlation $R_{12}$,  and the turbulence production term $\text{Prod}_\text{k}$ are extracted and shown in Fig.~\ref{fig:intro2}(c): 
\begin{equation}
\begin{aligned}
   R_{12}=\overline{\rho u_1'' u_2''}, \qquad
   \text{Prod}_\text{k}=-\overline{\rho u_i'' u_j''}\frac{\partial{\overline{u_i}}}{\partial{x_j}},
\end{aligned}
\end{equation}
where $''$ means the fluctuating part of velocity and the overbar represents the time averaging; $R_{12}$ represents momentum transport associated with large-scale unsteady flow motion rather than fully developed turbulent Reynolds stress. 
The purpose of this analysis is not to assess turbulence statistics, but to establish that the present flow field is characterized by substantial unsteady momentum transport typical of turbulence-affected jet impingement flows.
The spatial distribution of $R_{12}$ highlights regions of pronounced unsteadiness, including the jet exits, Mach-disk regions, and outer shear layers. The turbulence production term $\text{Prod}_k$ further identifies the precise locations of turbulence generation, which closely coincide with regions of elevated $\mu_r$ in Fig.~\ref{fig:intro}(c), indicating that the SST model captures the spatial trend of turbulent activity.

The cut plane at $x_2=-0.3$~m shown in Fig.~\ref{fig:intro2}(d) proves  the asynchronous variations of the NS stress, $R_{12}$, and the Boltzmann stress component. The NS stress undergoes multiple zero crossings, while the Boltzmann contribution remains finite. Direct inspection of instantaneous fields confirms that the same stress-degeneracy identified in the time-averaged results is also present at the instantaneous level. Time averaging therefore does not introduce this mechanism, but merely clarifies its spatial organization.


As a consequence of the degeneracy of the NS constitutive relation, the surface stress and heat flux predicted by NS-based direct numerical simulations systematically deviate from the Boltzmann solution, see Fig.~\ref{fig:intro}(b,d). By the way, the GSIS-SST solution captures the correct magnitude and spatial trends, albeit with moderate smoothing due to eddy-viscosity diffusion.

\textbf{Conclusions}---Based on the direct numerical simulation of the Boltzmann equation, we have identified that weak rarefaction effects can be developed in dual-jet impingement on solid surfaces. Although the rarefaction stresses remain small in absolute magnitude, they become dynamically decisive within localized shear layers where the leading-order NS constitutive contribution has weakened or has even undergone sign reversal. In such regions, the regular perturbative structure underlying continuum transport has broken down, allowing formally higher-order kinetic effects to have governed the local stress balance and the associated surface loads. This mechanism has explained why NS equations have substantially mispredicted shear stress and heat flux without exhibiting globally strong rarefaction signatures. 
More broadly, our findings have demonstrated that turbulent flows traditionally regarded as safely within the continuum regime may harbor localized breakdowns of constitutive validity, thereby addressing the long-standing question  regarding whether rarefaction effects in turbulence can significantly influence the flow field \citep{Tennekes1972first}. This finding has direct implications for aerodynamic and thermal predictions in extreme aerospace environments.

This work was supported by the National Natural Science Foundation of China (Grant No. 12450002). Special thanks are given to the Center for Computational Science and Engineering at the Southern University of Science and Technology.

\bibliography{ref}

\end{document}